\documentstyle[fleqn,twoside]{article}

\title{Evolving Algebras and Partial Evaluation%
\thanks{In {\em Proceedings of IFIP Congress 94 -- Volume 1\/}, eds.
B. Pehrson and I. Simon, Elsevier, 1994.}} 
\author{Yuri Gurevich\footnotemark[2] and James K. Huggins%
\thanks{Partially supported by ONR grant 
N00014-91-J-1861 and NSF grant CCR-92-04742.  
EECS Department, University of Michigan,
Ann Arbor, MI, 48109-2122, USA. gurevich@umich.edu,
huggins@umich.edu}}

\newif\ifEAfigs
\EAfigstrue

\newif\ifRuleLines
\RuleLinestrue

\newlength{\RuleWidth}
\newcommand{\BeginRule}{
\ifEAfigs \begin{figure}[htbp] 
\fi
\begin{center}
\ifRuleLines \rule{\RuleWidth}{.01in} \\ \fi
\begin{minipage}[t]{\RuleWidth}
\begin{em}
\begin{tabbing}
mmm\=mmm\=mmm\=mmm\=mmm\=mmm\=mmm\=mmm\=mmm\=mmm\=\kill
}

\newcommand{\EndRule}[2]{
\end{tabbing}
\end{em}
\end{minipage}
\ifRuleLines \rule{\RuleWidth}{.01in} \fi
\end{center}
\ifEAfigs  \caption{\label{#1} #2}
          \end{figure}  
\fi
}

\newcommand{\NoFigRules}{\EAfigsfalse}

\newcommand{\NoRuleLines}{\RuleLinesfalse}

\newcommand{\If}{{\bf if\ }}
\newcommand{\Then}{{\bf then\ }}
\newcommand{\Else}{{\bf else\ }}
\newcommand{\Elseif}{{\bf elseif\ }}
\newcommand{\Endif}{{\bf endif\ }}

\NoFigRules
\NoRuleLines

\begin{document}
\maketitle

\begin{tabbing}
Keyword Codes: D.2.2; D.2.m; F.3.2\\
Key\=words: Software Engineering, Tools and Techniques;\\
  \>Software Engineering, Miscellaneous;\\
  \>Logics and Meanings of Programs, Semantics of Programming
Languages
\end{tabbing}

\begin{abstract}
We describe an automated partial evaluator for evolving algebras
implemented at the University of Michigan.
\end{abstract}
\section{Introduction to Sequential Evolving Algebras}

A fuller discussion of evolving algebras (or {\em ealgebras\/}) can be
found in \cite{lipari}; to make this paper self-contained, we recall
briefly the main concepts.

A sequential ealgebra $\cal A$ is an abstract machine.  The {\em
signature\/} of $\cal A$ is a (finite) collection of function names,
each of a fixed arity.  A state of $\cal A$ is a set, the
{\em superuniverse\/}, together with interpretations of the function
names in the signature.  These interpretations are called {\em basic
functions\/} of the state.  The superuniverse does not change as $\cal
A$ evolves; the basic functions may.

Formally, a basic function of arity $r$ ({\it i.e.\/} the interpretation
of a function name of arity $r$) is an $r$-ary operation on the
superuniverse.  (We often use basic functions with $r=0$; such basic
functions will be called {\em distinguished elements\/}.)  But functions
naturally arising in applications may be defined only on a part of the
superuniverse.  Such partial functions are represented by total
functions in the following manner.

The superuniverse contains distinct elements {\em true, false,
undef\/} which allow us to deal with relations (viewed as binary
functions with values {\em true\/} or {\em false\/}) and partial
functions (where $f(\overline{a}) =$ {\em undef\/} means $f$ is
undefined at the tuple $\overline{a}$). These three elements are {\em
logical constants\/}.  Their names do not appear in the signature;
this is similar to the situation in first-order logic with equality
where equality is a logical constant and the sign of equality does not
appear in the signature.  In fact, we use equality as a logical
constant as well.

A {\em universe\/} $U$ is a special type of basic function: a
unary relation usually identified with the set $\{x: U(x)\}$.  The
universe $Bool = \{true, false\}$ is another logical constant.  When
we speak about a function $f$ from a universe $U$ to a universe
$V$, we mean that formally $f$ is a unary operation on the
superuniverse such that $f(a)\in V$ for all $a\in U$ and $f(a)=$ {\em 
undef\/} otherwise.  We use self-explanatory notations like 
$f: U \rightarrow V$, $f: U_1 \times U_2 \rightarrow V$, 
and $f: V$.  The last means that the distinguished element $f$
belongs to $V$.

In principle, a program of $\cal A$ is a finite collection of
transition rules of the form
\BeginRule
\If $t_0$ \Then f($t_1$, $\ldots$, $t_r$) := $t_{r+1}$ \Endif \`(1)
\EndRule{}{}
where $t_0$, $f(t_1, \ldots, t_r)$, and $t_{r+1}$ are closed terms
({\it i.e.\/} terms containing no free variables) in the signature of
$\cal A$.  An example of such a term is $g(h_1,h_2)$ where $g$ is
binary and $h_1$ and $h_2$ are zero-ary.  The meaning of the rule
shown above is this: Evaluate all the terms $t_i$ in the given state;
if $t_0$ evaluates to {\em true\/} then change the value of the basic
function $f$ at the value of the tuple $(t_1,..,t_r)$ to the value of
$t_{r+1}$, otherwise do nothing.

In fact, rules are defined in a slightly more liberal way;  
if $k$ is a natural number, \(b_0,\ldots,b_k\) are terms and
\(C_0,\ldots,C_{k}\) are sets of rules then the following
is a rule:
\BeginRule
\If $b_0$ \Then $C_0$\\
\Elseif $b_1$ \Then $C_1$\\
\ \ \vdots \`(2)\\
\Elseif $b_k$ \Then $C_k$\\
\Endif
\EndRule{}{}
In the case that $b_k = true$, the last line may be abbreviated by
``{\em \bf else} $C_k$''.  

Since the $C_i$ are sets of rules, nested
transition rules are allowed (and occur frequently). 

A program is a set of rules.  It is easy to transform a program to an
equivalent program comprising only rules of the stricter form (1).  We
use rules of the more liberal form (2), as well as macros (textual
abbreviations), for brevity.  

How does $\cal A$ evolve from one state to another?  In a given state,
the demon (or interpreter) evaluates all the relevant terms and then
makes all the necessary updates.  If several updates contradict each
other (trying to assign different values to the same basic function at
the same place), then the demon chooses nondeterministically one of
those updates to execute.

We call a function (name) $f$ {\em dynamic\/} if an assignment of the
form $f(t_1,\ldots,t_r) := t_0$ appears anywhere in the transition
rules.  Functions which are not dynamic are called {\em static\/}.  To
allow our algebras to interact conveniently with the outside world, we
also make use of {\em external\/} functions within our algebra.
External functions are syntactically static (that is, never changed by
rules), but have their values determined by a dynamic oracle.  Thus, an
external function may have different values for the same arguments as
the algebra evolves.

\subsection{Why Partial Evaluation and Evolving Algebras?}

One of the main application areas of ealgebras has been programming
language semantics.  One may view an ealgebra $A$ for a language $L$
as an abstract machine which acts as an interpreter for $L$.  With an
$L$-program $p$ as an input, $A$ gives semantics for $p$.  However,
$A$ may be large; for many programs, an ealgebra tailored directly to
$p$ is clearer than $A$.  Partial evaluation provides an automated
means for tailoring an ealgebra for an $L$-program $p$ by specializing
an ealgebra interpreter for $L$ with respect to $p$.  These tailored
ealgebras may not be as good as hand-tailored ones, but they may
provide a useful beginning for tailored ealgebras.  Of course, this is
only one use of a partial evaluator for ealgebras.

\section{Partial Evaluation Techniques}

Suppose one has a program $p$ and knows a portion of its input ahead
of time.  Can one take advantage of this information to transform $p$
into a more efficient program?  {\em Partial evaluation\/} is the
process of transforming such a program $p$ into a program
$p'$ which, when supplied with the remainder of $p$'s input, has the
same behavior as $p$.

Our partial evaluator follows the ``mix'' methodology (described in
more detail in \cite{jgs}) and has three phases: binding-time
analysis, polyvariant mixed computation, and post-processing
optimizations.  We describe each of these phases below.

\subsection{Binding-Time Analysis}

Initially, the partial evaluator is given the names of the basic 
functions of the ealgebra which will be known ahead of time.  During {\em
binding-time analysis\/}, the partial evaluator determines 
which basic functions can be pre-computed in the next phase.  
This process is called binding-time analysis because it determines at
what time the value(s) of a basic function can be determined ({\em
i.e.\/} bound to known values).

The input to this phase is a division of the basic functions which
supply input to the ealgebra into two sets: {\em positive\/} functions
whose values will be known ahead of time, and {\em negative\/} functions
whose values will not be known until later.  The partial evaluator
proceeds to classify all basic functions (including those not
initially marked by the user) as positive or negative.  (\cite{jgs}
use the terms ``static'' and ``dynamic'' to refer to these types of
values; these terms have different meanings within the ealgebra paradigm.)

In the current implementation, the following algorithm is used to
classify a function $f$ as positive or negative:
\begin{itemize}
\item If $f$ is syntactically static (that is, not updated by any
transition rule), $f$ remains as classified initially by the user.
\item If an update $f(\bar{t}) := t_0$ exists in $p$ such that $\bar{t}$ or
$t_0$ references a negative function, $f$ is negative.  (Note that
even if $f$ was declared as positive by the user, $f$ may still depend
on other negative functions and must be classified as negative.)
\item If for all updates $f(\bar{t}) := t_0$ in $p$,
every function referenced in $\bar{t}$ and $t_0$ is
positive, and $f$ is not already negative, $f$ is positive.
\end{itemize}
This classification algorithm is repeatedly applied until a
fixed-point is reached.  Any remaining unclassified functions are
classified as negative and the algorithm is repeated to ensure
consistency.

An interesting problem which this algorithm does not handle is the
problem of circular dependencies.  
A basic function $f$ is {\em
self-referential\/} if some update $f(\bar{t}) := t_0$ within the
ealgebra being specialized contains a reference to $f$ within
$\bar{t}$ or $t_0$.  The above algorithm classifiess every
self-referential function as negative.  Often this is appropriate, as
some self-referential functions can grow unboundedly.  But at times,
classifying such functions as positive is also appropriate.  Consider
the following program:

\BeginRule
\If Num $>$ 0 \Then Num := Num + 1 \Endif\\
\If MyList $\not=$ Nil \Then MyList := Tail(MyList) \Endif
\EndRule{}{}

Suppose the initial values of {\em Num\/} and {\em MyList\/} are
known.  {\em Num\/} should not be classified as a positive function,
since it would lead the specializer in the next stage into an infinite
loop, as larger and larger values of {\em Num\/} would be computed as
positive information.  On the other hand, there is no problem with
classifying {\em MyList\/} as positive, since {\em MyList\/} will
eventually be reduced to {\em Nil\/} and remain at that value forever.
An addition to the algorithm presented above properly classifies
self-referential functions as positive if they are dependent only upon
themselves in a bounded manner (as seen here).  

The problem of circular dependencies is much more general that the
problem of self-reference; it may be that several functions form a
mutual dependency cycle.  We intend to incorporate a more
sophisticated algorithm for binding-time analysis based on an
examination of the dependency graph formed by the basic functions of
the algebra.  In the future, we hope to extend this analysis to parts
of basic functions; it may be that $f(\bar{t})$ could be classified as
positive for certain tuples $\bar{t}$ but not for others.

\subsection{Polyvariant Mixed Computation} 
After binding time analysis, the partial evaluator begins the process
of specializing the input program, executing rules which depend
only on positive information (that is, functions classified as
positive by our binding-time analysis) and generating code for rules
which depend on negative information.  The process is called {\em
polyvariant mixed computation}: ``mixed'' because the processes of
executing positive rules and generating code for negative rules is
interleaved, and ``polyvariant'' because the entire program is
considered multiple times for different sets of positive information.

The signature $\tau$ of the algebra has been divided into two
components during binding time analysis: a positive signature
$\tau_+$ and a negative signature $\tau_-$. This leads to a
corresponding division of states (or structures) $S$ into structures
$S_+$ and $S_-$.  The partial evaluator creates an ealgebra with
signature $\tau_- \cup
\{K\}$, where $K$ is a nullary function which will be
used to hold the positive (or ``known'') information formerly stored
by functions in $\tau_+$.

From a given positive state $S_+$, the partial evaluator produces rules
of the form
\BeginRule
\If K = $S_+$ \Then rules \Endif
\EndRule{}{}
where {\em rules\/} is a specialized version of the rules of the
entire input program with respect to $S_+$, along with an assignment
to {\em K\/}.  Call a transition rule of this form a {\em K-rule\/},
whose {\em guard\/} is {\em (K = $S_+$)\/} and whose {\em body\/} is
{\em (rules)\/}.  Not that no two K-rules produced by our partial
evaluator will have the same guard.  We recursively describe how
transition rules are speciailized with respect to a given positive
state $S_+$ below.

An expression is specialized with respect to $S_+$ by
substituting all known values of functions in $S_+$ into the given
expression, simplifying when possible.

An update $f(\bar{t}) := t_0$ is specialized with respect to 
$S_+$ as follows.  If $f \in \tau_+$, no rule is
generated.  Instead, the change to the positive function $f$ is noted
internally in order to generate the correct assignment to $K$.  (Note
that in this case, all functions named in $\bar{t}$ and $t_0$ are
positive as a result of our binding time analysis.)  Otherwise, an
assignment to $f$ is generated, with the values of $t_0$ and $\bar{t}$
specialized as much as possible using the information in $S_+$.

A set of rules is specialized with respect to $S_+$ by
specializing each rule and combining the information needed to create
a single assignment to $K$.

A guarded rule ``{\em {\bf if} guard {\bf then} $R_1$ {\bf else} $R_2$
{\bf endif}}'' is specialized with respect to 
$S_+$ as follows.  If all functions in {\em guard\/} are positive, the
result is the specialization of $R_1$ or $R_2$, depending on whether
the value of {\em guard\/} is true or false in $S_+$.  Otherwise, an
{\em \bf if} statement is generated, with {\em guard\/}, $R_1$, and
$R_2$ specialized as above.  A guarded rule containing {\em \bf
elseif} clauses is converted to an equivalent form without {\em \bf
elseif} clauses and specialized as above.

\subsection{Optimization}  

The above transformations create a specialized version of the original
program.  Often, this specialized version contains many unneeded
rules, such as:
\BeginRule
\If K = foo \Then K := bar \Endif
\EndRule{}{}
Such a K-rule can be deleted; just replace all
references to {\em foo\/} in the program with references to
{\em bar}.  The partial evaluator performs several such optimizations
on the specialized program:
\begin{itemize}
\item Eliminating terms which serve only as aliases for other terms or
constants. 
\item Combining K-rules with identical bodies.
\item Combining K-rules which are executed consecutively but whose
bodies have independent updates that could be executed simultaneously
without altering the meaning of the program.
\item Eliminating K-rules which will never be executed.
\end{itemize}
These optimizations generate code which is equivalent, but textually
shorter and usually requires fewer moves to execute.  Of course, one
pays a small price in time in order to generate these optimizations.

\subsection{Results}

It is important that a partial evaluator actually perform useful work.
Kleene's s-m-n theorem shows that partial evaluators can in principle
be constructed; his proof shows that such evaluators may not
necessarily produce output that is more efficient than the original.
One can specialize an ealgebra, for example, by creating two K-rules:
one which initializes the functions in $S_+$ to their initial values
and one which has the original unspecialized program as its body.
This ``specialized'' ealgebra has the same behavior as the original,
but is hardly more useful than the original algebra.

\cite{jgs} suggest a standard for evaluating partial evaluators.
Consider a self-interpreter for a language: that is, an interpreter
for a language $L$ written itself in $L$.  (McCarthy's original
description of a LISP interpreter written itself in LISP is such an
interpreter.)  Specializing such an interpreter with respect to an
$L$-program $p$ should yield a version of $p$, of size comparable to
$p$, as output.  That is, the overhead involved in interpreting a
language (deciding which command is to be executed, incrementing a
program counter, {\em etc.\/}) should be removed by a good partial
evaluator.  Our partial evaluator seems to approach this standard when
run on small programs, though more detailed testing is needed.

\section{An Example}

Consider the C function {\tt strcpy}:

\begin{center}
{\tt {void strcpy (char *s, char *t) \verb+{+ 
                while (*s++ = *t++) ; \verb+}+ } }
\end{center}


This function copies a string from the memory location indicated by 
$t$ to the memory location indicated by $s$.  It is admittedly
cryptic.

In \cite{c}, we presented an ealgebra interpreter for the C
programming language.  As a test, we ran our partial evaluator on our
algebra for C, specializing it with respect to {\tt strcpy()}.  The
result, with most of the functions renamed, appears below.

\BeginRule
\If K = ``init'' \Then
   CopyFrom := {\tt t}, CopyTo := {\tt s}, K := ``first-update''
\Endif\\
\If K = ``first-update'' \Then\\
   \>TmpFrom := CopyFrom, TmpTo := CopyTo\\
   \>CopyFrom := CopyFrom + 1, CopyTo := CopyTo + 1, K := ``loop''\\
\Endif\\
\If K = ``loop'' \Then\\
   \>\If Memory(TmpFrom) $\not=$ 0 \Then\\
   \>   \>Memory(TmpTo) := Memory(TmpFrom)\\
   \>   \>TmpFrom := CopyFrom,  TmpTo := CopyTo\\
   \>   \>CopyFrom := CopyFrom + 1, CopyTo := CopyTo + 1, K := ``loop''\\
   \>\Else Memory(TmpTo) := Memory(TmpFrom), K := ``done''\\
   \>\Endif\\
\Endif
\EndRule{}{}

This algebra is considerably smaller than the entire ealgebra for C,
and hopefully is more easily understandable than the original C code.
It is not optimal: for example, {\em CopyFrom\/} could be replaced by
{\tt t}, since {\tt t} is never used after the initial state.  It
does, however, exhibit the behavior of {\tt strcpy()} more directly
than the entire ealgebra for C given {\tt strcpy} as input.

(For those familiar with \cite{c}, the input functions
initially specified as positive were {\em CurTask\/}, {\em
TaskType\/}, {\em NextTask\/}, {\em LeftTask\/}, {\em RightTask\/},
{\em TrueTask\/}, {\em FalseTask\/}, {\em Decl\/}, {\em Which\-Child\/},
and {\em ChooseChild\/}, assuming that {\em ChooseChild\/} always
moves to the left first.)

\end{document}